\documentclass[a4paper,conference]{IEEEtran}
\usepackage{multirow}

\ifCLASSINFOpdf
  \usepackage[pdftex]{graphicx}
\else
\fi
\usepackage{amsmath}

\begin{document}

%
\title{Transfer Learning Through Weighted Loss Function and Group Normalization for Vessel Segmentation from Retinal Images}



%
\author{\IEEEauthorblockN{
Abdullah Sarhan\IEEEauthorrefmark{1}\IEEEauthorrefmark{5},
Jon Rokne\IEEEauthorrefmark{1},
Reda Alhajj\IEEEauthorrefmark{1}\IEEEauthorrefmark{2}\IEEEauthorrefmark{3} and
Andrew Crichton\IEEEauthorrefmark{4}}
\IEEEauthorblockA{\IEEEauthorrefmark{1}Department of Computer Science, University of Calgary,  Alberta, Canada}

\IEEEauthorblockA{\IEEEauthorrefmark{2}Department of Computer Engineering, Istanbul Medipol University, Istanbul, Turkey}
\IEEEauthorblockA{\IEEEauthorrefmark{3}Department of Health Informatics, University of Southern Denmark, Odense, Denmark}
\IEEEauthorblockA{\IEEEauthorrefmark{4}Department of Ophthalmology and Visual Sciences, University of Calgary, Alberta, Canada}
\IEEEauthorblockA{\IEEEauthorrefmark{5}Corresponding author, email:asarhan@ucalgary.ca}
}


\maketitle

\begin{abstract}
The vascular structure of blood vessels is important in diagnosing retinal conditions such as glaucoma and diabetic retinopathy. Accurate segmentation of these vessels can help in detecting retinal objects such as the optic disc and optic cup and hence determine if there are damages to these areas. Moreover, the structure of the vessels can help in diagnosing glaucoma. The rapid development of digital imaging and computer-vision techniques has increased the potential for developing approaches for segmenting retinal vessels. In this paper, we propose an approach for segmenting retinal vessels that uses deep learning along with transfer learning. We adapted the U-Net structure to use a customized InceptionV3 as the encoder and used multiple skip connections to form the decoder. Moreover, we used a weighted loss function to handle the issue of class imbalance in retinal images. Furthermore, we contributed a new dataset to this field. We tested our approach on six publicly available datasets and a newly created dataset. We achieved an average accuracy of 95.60\% and a Dice  coefficient of 80.98\%. The results obtained from comprehensive experiments demonstrate the robustness of our approach to the segmentation of blood vessels in retinal images obtained from different sources. Our approach results in greater segmentation accuracy than other approaches.
\end{abstract}


%
\IEEEpeerreviewmaketitle

\section{Introduction}

Sight is one of the most important senses for humans. It allows us to visualize and explore our surroundings and any conditions that reduce the vision impacts our ability to interact with the surroundings. Several degenerative ocular conditions have been identified, such as glaucoma and diabetic retinopathy. These conditions threaten our sense of sight by causing irreversible visual-field loss \cite{sarhan2019glaucoma}. Glaucoma is is a condition that is the world’s second most prominent cause of irreversible vision loss after cataracts, accounting for 12\% of cases of blindness worldwide \cite{fu2017segmentation}.

The structure of retinal blood vessels serves an important role in the diagnosis of retinal conditions such as glaucoma \cite{sarhan2019glaucoma}. For instance, one of the indicators of glaucoma is deviation from the inferior, superior, nasal, and temporal (ISNT) rule, which states that the inferior neuro-retinal rim should be thickest, followed by the superior rim, the nasal, and the temporal. Additionally, changes within the optic disc, such as the displacement of blood vessels can be used to help determine if glaucoma is present and if the disease has progressed \cite{issac2015adaptive}.

With the advancements in the field of image processing and deep learning, it is possible to develop an automated approach for segmenting vessels with reliable precision. Such a method could help in analyzing these vessels when diagnosing retinal conditions such as glaucoma. The process would involve two challenges: first, segmenting the vessels; and second, analyzing the structure by extracting features that are highly correlated with a specific retinal condition.

In this study, we focus on solving the first challenge by proposing a deep-learning approach for segmenting the vessels in retinal images. Our model is based on the U-Net architecture and uses the InceptionV3 \cite{szegedy2016rethinking} model as the encoder. Given the challenges related to having insufficient annotated retina vessels datasets for deep learning, we adopted the idea of using transfer leaning and image augmentation. Instead of using random weights to initialize our model, we use weights trained on millions of images for semantic segmentation from the ImageNet dataset, which we then fine-tune to match the object to be segmented (i.e., retinal vessels). To handle the issue of imbalanced classes, we use a customized loss function that tends to penalize incorrect classifications of pixels related to vessels more than incorrect classifications of background pixels. 

The proposed model takes the whole retinal image as input and then segments the vessels in a short time (1 second). Instead of using batch normalization on a small number of batches, we used group normalization. To demonstrate the robustness of our proposed approach in segmenting the retinal vessels from images with various resolutions, we tested our approach on six publicly available datasets and one private dataset that we created. Our contributions can be enumerated as follows: (1) We propose a U-Net based deep learning model for vessel segmentation that utilizes a customized InceptionV3 model as an encoder. (2) We demonstrate the effectiveness of using TL and IA for limited data. (3) We handle the issue of imbalanced image classes which may lead to inaccurate results by adopting a weighted loss function. (4) We contribute a new retinal image dataset for vessels segmentation (ORVS).

\begin{figure*}[htb!]
\centering
\includegraphics[width=\linewidth,height=8.4cm,keepaspectratio]{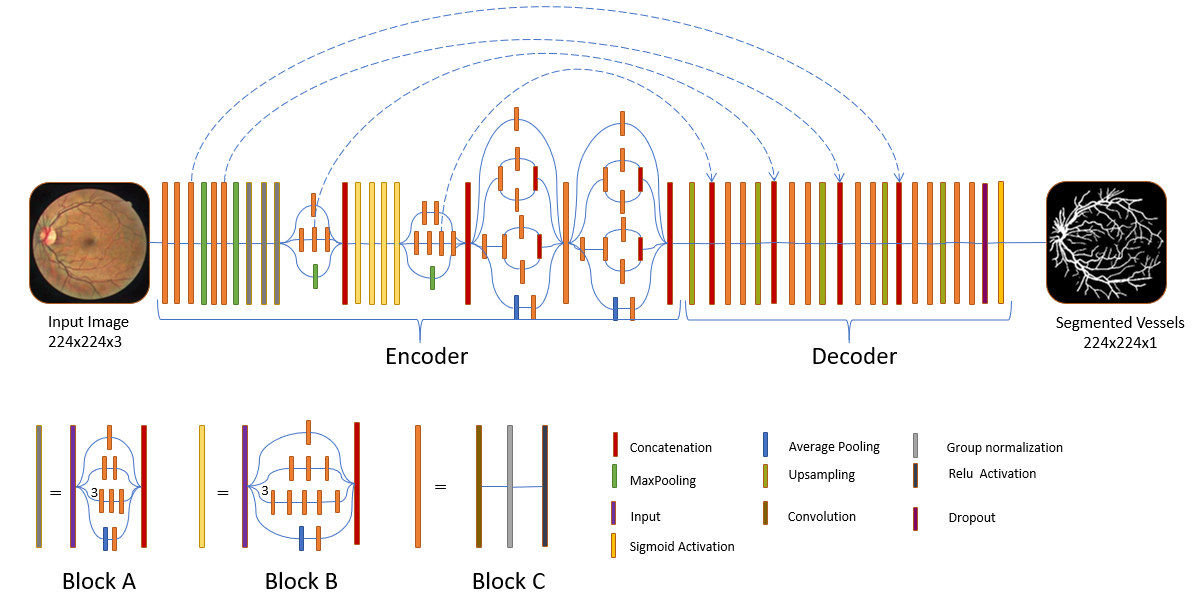}
\caption{Architecture of the  model adopted in this study.}
\label{fig:InceptionV3UnetModel}
\end{figure*}

\section{Related Work}
\label{Related Work}

Several approaches have been proposed for retinal vessel segmentation. These approaches can generally be categorized into unsupervised and supervised approaches. The literature reviews conducted by \cite{akbar2019automated} and \cite{sarhan2019glaucoma} discuss more than a hundred  approaches for vessel segmentation that are either supervised or unsupervised.

Unsupervised approaches segment the vessels using predefined sets of rules. If a pixel passes a set of conditions, it is considered to be related to a vessel; otherwise it is considered to be unrelated. Such approaches use use variety of techniques such as morphological operations \cite{mendonca2006segmentation}; matched filters \cite{khan2016automatic,bankhead2012fast}; multi scale \cite{nguyen2013effective}; adaptive thresholding \cite{roychowdhury2015iterative,zhao2014retinal}. They may also be model-based \cite{memari2019retinal}. Unsupervised approaches tend to offer faster segmentation than supervised ones; however, they tend to fail when exposed to thin vessels or images with low contrast.

The other category of approaches to retinal vessel segmentation is supervised approaches. In such approaches, a pixel matrix $I$ is associated with each retinal image, indicating the pixels that belong either to vessel pr to background. $I_{xy}$ represents a pixel at location $(x,y)$ in the retinal image. This pixel has a value of 1 if it belongs to a vessel and 0 if it is a background pixel. The classifier uses these labeled images and the actual retinal image to produce a new image with the same dimensions in which each pixel has a probability between 0 and 1 inclusive of belonging to a vessel. The closer the value is to 1, the higher the model’s confidence is that the pixel belongs to a vessel.

Recently, advancements in the field of deep learning have made it possible to use deep learning-based classifiers in medical-image analysis. Such approaches exhibit superior performance over the hand-crafted ones \cite{sarhan2019glaucoma,shen2017deep}. Various deep-learning architectures have been proposed for vessel segmentation. For instance, in \cite{xiao2018weighted}, U-Net architectures were used with residual blocks and binary cross-entropy as the loss functions. In this study they extracted patches from retinal images for the segmentation instead of performing the segmentation from the whole image. In analyzing the images, however, such models face the issue of imbalanced classes. Another model architecture used is fully connected layers, which was adopted by \cite{oliveira2018retinal} and \cite{hu2018retinal}. Neither handled the issue of imbalanced classes, however. Moreover, they tested their approach  with a limited number of images.

Vessel segmentation poses several challenges, such as limited availability of datasets and the imbalance between the number of pixels related to the vessels and those related to other classes. Another challenge is related to the number of images used per batch when training the deep-learning model leading to increasing the batch error and hence inaccurate statistics. In this paper, we show that using image augmentation and transfer learning with a weighted loss function and customized U-Net model can help in achieving high precision when segmenting the retinal vessels. We also created a new dataset that can help researchers working in this field.

\section{Proposed Method}
The goal of this study is to segment the vessels in a retinal image. To achieve this, we propose a deep-learning model with the same architecture as the U-Net model \cite{ronneberger2015u} and use a customized InceptionV3 model as the encoder. In this section we discuss the architecture of our model, the prepossessing steps, data augmentation, and model initialization.

\subsection{Network Architecture}
Instead of creating a new architecture, we adopted the U-Net architecture, which consists of an encoder and a decoder. In the encoder, pooling is applied to summarize neighboring features and create high-level representations. In the decoder, feature maps are reconstructed, combining those from the encoder through skip connections, with those belonging to low-level scales.

We adopted the InceptionV3 model as the encoder, customized the normalization, and reduced the convolutional layers. The architecture of our model is presented in Fig. \ref{fig:InceptionV3UnetModel}. The original architecture of InceptionV3 requires four instances of block C to be presented in block A (marked as as 3 in Block A in Fig.  \ref{fig:InceptionV3UnetModel}) but we used only three. Another change is that, in the original architecture, block B is required to have four instances of block C, but we used five instead (marked as 3 in Block B in Fig.  \ref{fig:InceptionV3UnetModel}). We also removed the auxiliary classifier. The final concatenation layer and other layers as shown in Fig. \ref{fig:InceptionV3UnetModel} were used to construct our encoder. A drop-out was added at the end with a factor of 0.3, and a sigmoid activation was used. The feature map for each convolutional layer uses the ReLU activation method, which applies Eq. \ref{equ:relU} to each parameter from the layer, thereby removing all negative pixel values.

\begin{equation}
\label{equ:relU}
f(x)={max(0,x)}
\end{equation}

The model is initially fed with a preprocessed image of size $224\times224$. Three channels are used, not only the green channel as used by other approaches \cite{sarhan2019glaucoma}. Skip connections were used in several places in the architecture. In Fig. \ref{fig:InceptionV3UnetModel}, the first skip connection is at the third occurrence of block C. Other skip connections were then added to form the decoder. The selection of blocks to include in the encoder was based on a set of experiments we conducted. After the skip layers had been selected, we performed upsampling and concatenation so that the output image will have a $224\times224$ dimensions. This was used to shorten the distance between the earlier and later layers. Short connections from early to later layers are useful in preserving high-level information about the positioning of the vessels. This is in contrast to the low-level pixel-based information that is transferred across the long pipeline of the architecture in a combination of convolutional and max-pooling or upsampling layers. High-level information tends to be lost as the image is down-sampled and the shape and structure of the image is changed. We maintain this information by using connections to earlier layers in the model.

\subsection{Preprocessing}
Retinal images obtained in less than ideal imaging environments tend to be of poor quality. In such images, the appearance of the blood vessels may be affected due to fluctuating pixels intensities or the addition of noise to the image. To handle these issues, we applied some image-processing techniques. These techniques were applied to every image before augmentation.

\begin{figure}[htb!]
\centering
\includegraphics[width=\linewidth,height=3cm,keepaspectratio]{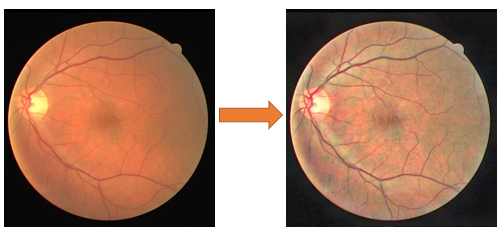}
\caption{Retinal Images (Left) after doing some enhancement (Right).}
\label{fig:PreprocessingImages}
\end{figure}

To deal with the issue of contrast variation, we use the contrast-limited adaptive histogram equalization CLAHE \cite{pizer1987adaptive} for each channel. This technique can enhance the contrast by stretching the grey level values of low contrast images. Gamma correction is then used to adjust the brightness. To smooth the edges of vessels, we apply a median filter, which computes the median of all the pixels under the kernel window (in our case it is $5\times5$), and the central pixel is replaced with this median value. All images are then resized to $224\times224$ and normalized to ensure every pixel value is between 0 and 1. Fig. \ref{fig:PreprocessingImages} shows an original image before (left) and after (right) these techniques have been applied.

\subsection{Group Normalization layer}
In large models such as the one in the present study, normalization, including batch normalization, is important for training and convergence. Batch normalization works well when the number of images per batch is high (greater than 32); however, the number of errors in batch normalization increases as the number of images decreases, leading to inaccurate batch statistics. To overcome this issue, we use group normalization, which has proved to be effective in such situations \cite{wu2018group}. The group normalization divides the channels into groups and then calculates the mean and variance among the groups. Hence, group normalization is independent of batch size.

In training we used a batch size of two, because that was the maximum capacity of our machine memory. Group normalization organizes the channels into groups and calculates the mean and variance along the height and width axis and along a group channels  using Eq. \ref{equ:GN} where $\mu$ and $\sigma$ are computer over a set of pixels defined by $S_i$ where $S_i$ is defined using Eq. \ref{equ:SI}. In Eq. \ref{equ:SI}, $G$ represents groups which is in our case is set to 16,  C/G represents the number of channels per groups, and the $\lfloor.\rfloor$ represents the floor operation.

\begin{equation}
\label{equ:GN}
\mu_i=\frac{1}{m}\sum_{k\in S_i} x_k , \sigma_i=\sqrt{\frac{1}{m}\sum_{k\in S_i} (x_k-\mu_i)^2+\epsilon}
\end{equation}

\begin{equation}
\label{equ:SI}
S_i=\lbrace k | k_N=i_N, \Big\lfloor\frac{k_C}{C/G}\Big\rfloor,\Big\lfloor\frac{i_C}{C/G}\Big\rfloor \rbrace
\end{equation}

\subsection{Loss Function}
In order to check if the model had improved or not on the value returned from the loss function during the training of the network it was run using the validation data. At first the binary cross-entropy function (BCE), shown in Eq. \ref{equ:BCE} was adopted where the total number of pixels is denoted by $N$, and the pixels $y_i $ are labelled by 0 for background and 1 for the vessels. The probability that pixel $y_i $ belongs to the vessel is $p_d(y_i)$ and to the background  $p_b(y_i)$. It should be noted that during classification using the BCE both false positives and false negatives can be penalized when working on classifying vessels and background pixels.
\begin{equation}
\label{equ:BCE}
BCE=- \frac{1}{N}\sum_{i=1}^N y_i.\log (p(y_i)) +(1-y_i).\log(1-(p(y_i)))
\end{equation}
Using the BCE loss function will bias the output towards the background resulting in improper vessel segmentation.  That is, the output may suggest a misleading result of 90\% accuracy. To achieve precise vessel segmentation output therefore require additional processing. This problem was avoided using the Jaccard distance which measures the dissimilarity between two data sets. It is defined as:
\begin{equation}
\label{equ:JLoss}
L_{j}=1- \frac{|Y_d \cap \hat{Y_d}|}{|Y_d \cup \hat{Y_d}|}
= 1- \frac{\sum_{d\in Y_d} (1 \land \hat{y_d})}{|Y_d| + \sum_{b\in Y_b} (0 \lor \hat{y_b})}
\end{equation}

$Y_d$ represent the ground truth of the vessels, $Y_b$ the ground truth of background, $\hat{Y_d}$ the predicted vessel pixels and and $\hat{Y_b}$ represent the predicted background pixels. The cardinalities of the vessels $Y_d$ are $|Y_d|$ |and the cardinality background $|Y_b|$ is $|\hat{Y_b}|$ and $\hat{y_d}\in\hat{Y_d}$ and $\hat{y_b}\in\hat{Y_b}$. The values of $\hat{y_d}$ and $\hat{y_b}$ will always be between 0 and 1 since they are both probabilities. We can now approximate this loss function as shown in Eq. \ref{equ:JApproximated}. The model will then be updated by Eq. \ref{equ:JDerived} where $j$ represents the the $j$th pixel in the input image and $\hat{y_j}$ represents the predicted value for that pixel. 

\begin{equation}
\label{equ:JApproximated}
\tilde{L_j} = 1- \frac{\sum_{d\in Y_d} min(1, \hat{y_d})}{|Y_d| + \sum_{b\in Y_b} max(0,\hat{y_b})} \\
= 1- \frac{\sum_{d\in Y_d} \hat{y_d}}{|Y_d| + \sum_{b\in Y_b} \hat{y_b}}
\end{equation}
\begin{equation}
\label{equ:JDerived}
{L_j}{y_i} 
\begin{cases}-\frac{1}{|Y_d| + \sum_{b\in Y_b} \hat{y_b}} &  \hspace{0.5cm} for  \hspace{0.5cm} i \in Y_d\\
\\-\frac{\sum_{d\in Y_d} \hat{y_d}}{|Y_d| + \sum_{b\in Y_b} \hat{y_b}} & \hspace{0.5cm} for \hspace{0.5cm} i \in Y_b\end{cases}
\end{equation}
We are now able to balance the emphasis the model gives to the  vessel class and the background class using the Jaccard loss function and thus we combine BCE with Jaccard to optimize the results. When both are combined, the model converges faster than when using only the Jaccard still getting better results than either the  BCE or the Jaccard loss function by themselves. Our final loss function is therefore with $\beta_1$=0.75 and $\beta_2$=0.25:
\begin{equation}
\label{equ:FinalLoss}
L_f=\beta_1 \times BCE+ \beta_2 \times L_j
\end{equation}

\subsection{Transfer learning}
To handle the challenge faced in the field of medical imaging of not having enough datasets or large enough datasets to train a deep-learning model, we used an approach referred to as transfer learning. Such approaches can alleviate the issues caused by insufficient training data by using weights generated by training on millions of images \cite{pan2009survey}. In our study, we adopted the weights generated when training the InceptionV3 model on the ImageNet dataset which contains around 14 million labeled images. We thus provided a diverse set of images that the model had been exposed to.

By using transfer learning, we could reduce the problem of over-fitting caused when training on limited images and improve the overall performance of the model. Using the ImageNet weights, we initialized the weights of the encoder network component, and other layers were randomly initialized using a Gaussian distribution. We then trained our model using a mini-batch gradient to tune the weights of the whole network. When training, we realized when using transfer learning that the model converged faster than without transfer learning.

\subsection{Image Augmentation}
In total there were only 271 retinal images available for training, which is insufficient even when transfer learning is used. We thus augmented the dataset by applying a series of operations. These operations increased the number of images while preserving the characteristics of vessels. Hence, they allowed the model to be exposed to vessels of various appearances.

Unlike the approach in \cite{xiao2018weighted} we do not split the images into small patches. Doing so would require us to split the images into patches whenever we needed to segment the vessels, perform the segmentation and then re-assemble the images. Instead our approach maintains the appearance of vessels while increasing the number of images. We split each image into two halves vertically and two halves horizontally, producing a total of five images including the original. Each of these images is rotated by 90, 180, and 270 degrees, giving a total of 20 images $(5+5\times3)$. We then flipped both vertically and horizontally. Hence, from one image, we can obtain 60 images ($20+20\times2$), which increased our dataset from 271 images to 16,260 $(60\times271)$. This larger dataset improves the generalization performance of our model.

\begin{table*}[htb!]
\center
\caption {\label{table:Datasets}  Approaches for Retinal Vessels Segmentation.} 
\begin{tabular}{lccccc}
\hline
Dataset                                                    & \multicolumn{2}{c}{Images}                           & Dimensions                                                                                    & FOV                  & Machine                                                                                                  \\ \cline{2-3}
                                                           & \multicolumn{1}{l}{Train} & \multicolumn{1}{l}{Test} & \multicolumn{1}{l}{}                                                                          & \multicolumn{1}{l}{} & \multicolumn{1}{l}{}                                                                                     \\ \hline
ARIA\cite{bankhead2012fast}               & 120                 & 22                 & 768$\times$584                                                                                & 50                  & Canon CR5 non-mydriatic 3CCD                                                                             \\ \hline
CHASE\cite{fraz2012ensemble}            & 20                   & 8                 & 999$\times$960                                                                               & 30                  & Handheld Camera NM-200-D                                                                                 \\ \hline
\multirow{2}{*}{DR-Hagis\cite{holm2017dr}} & \multirow{2}{*}{34} & \multirow{2}{*}{6} & 3216$\times$2136, 4752$\times$3168                                                            & \multirow{2}{*}{45} & \multirow{2}{*}{\begin{tabular}[c]{@{}c@{}}Canon CR DGi, Topcon TRC NW6, \\ Topcon TRC NW8\end{tabular}} \\
                                                           &                     &                    & \begin{tabular}[c]{@{}c@{}}3456$\times$2304,2896$\times$1944,\\ 2816$\times$1880\end{tabular} &                     &                                                                                                          \\ \hline
DRIVE\cite{niemeijer2004comparative}      & 20                  & 20                 & 650$\times$500                                                                                & 35                  & TOPCON TRV-50                                                                                            \\ \hline
HRF\cite{odstrcilik2013retinal}           & 15                  & 30                 & 3504$\times$2336                                                                              & 60                  & Canon CF 60 UVI with Cannon EOS 20D                                                                      \\ \hline
STARE \cite{hoover2000locating}           & 20                  & 20                 & 700$\times$605                                                                                & 45                  & Topcon TRC NW6                                                                                           \\ \hline
ORVS                                                       & 42                  & 7                  & 1444$\times$1444                                                                              & 30                  & Zeiss, Viscuam 200                                                                                       \\ \hline
\end{tabular}
\end{table*}

\section{EXPERIMENTAL Analysis}
\label{Experimental Setup}
\subsection{Data sets}
\subsubsection{New dataset}
One of the challenges faced when working with retinal images is the limited availability of images. To help increase the number of images available for researchers working in this field, we introduce a new online retinal image for vessel segmentation (ORVS). The ORVS dataset has been newly established as a collaboration between the computer-science and visual-science departments at the University of Calgary.\footnote{https://tinyurl.com/y3kr7o4j} 

This  dataset contains 49 images (42 training and seven testing images) collected from a clinic in Calgary-Canada (check Table \ref{table:Datasets}). All images were acquired with a Zeiss Visucam 200 with 30 degrees field of view (FOV). The image size is 1444$\times$1444 with 24 bits per pixel. Images and are stored in JPEG format with low compression, which is common in ophthalmology practice. All images were manually traced by an expert who a has been working in the field of retinal-image analysis and went through training. The expert was asked to label all pixels belonging to retinal vessels. The Windows Paint 3D tool was used to manually label the images.

\subsubsection{Public datasets}
To evaluate our approach on images obtained from different FOVs and in various resolutions, we decided to use six  publicly available datasets: namely, the ARIA, CHASE, DR-Hagis, DRIVE, HRF, and STARE. Details about each dataset can be found in Table \ref{table:Datasets}. Using these datasets allows us to compare our approach with other state-of-the-art approaches that have evaluated their method with respect to some of these datasets. 

The DRIVE and CHASE datasets contain annotations made by multiple experts. In this study, we used the annotations made by the first observer as a ground truth. For the STARE dataset, we used the annotations made by Hoover. For the HRF dataset. we used the binary gold standard provided. As for the ARIA and DR-Hagis datasets, only one annotation was provided, and we used that for our training and testing datasets. The ground truth used for these studies is the   one commonly used, which allows us to make a fair comparison with other approaches

All datasets were divided into training, validating, and testing subsets. The training set was used for adjusting the weights while the validation set was used for saving the be model. These sets are selected randomly at each epoch. The testing dataset was used for evaluating the performance of our approach. For the DRIVE dataset, images were already split into 20 for training and validation and 20 for testing which was used by all approached to evaluate their model. The rest of the datasets do not have the same predefined splitting and hence we had to decide on how to split these images fairly so that we can compare with other approaches. For the STARE dataset, we use the "leave-one-out" method which is the common approach used when testing on this dataset \cite{yan2018joint,wang2019blood,orlando2016discriminatively}. In this method, we train the model on 19 images and test it on one. We repeat the same process until we test on all images and then we calculate the average. For the CHASE dataset, we adopt the approach in \cite{yan2018joint,wang2019blood} of which we selected the first 20 images for training and validation and last 8 for testing. For the HRF dataset, we also adopted the same approach used by \cite{yan2018joint} of which we take the first image 5 images from each category (healthy, diabetic retinopathy, and glaucoma) for training and validation and tested on all remaining images. As for DR-Hagis, ARIA, and ORVS we split by selecting randomly 15\% of the images to be as testing and the remaining for training and validation.

\begin{figure*}[htb!]
\centering
\includegraphics[width=\linewidth,keepaspectratio]{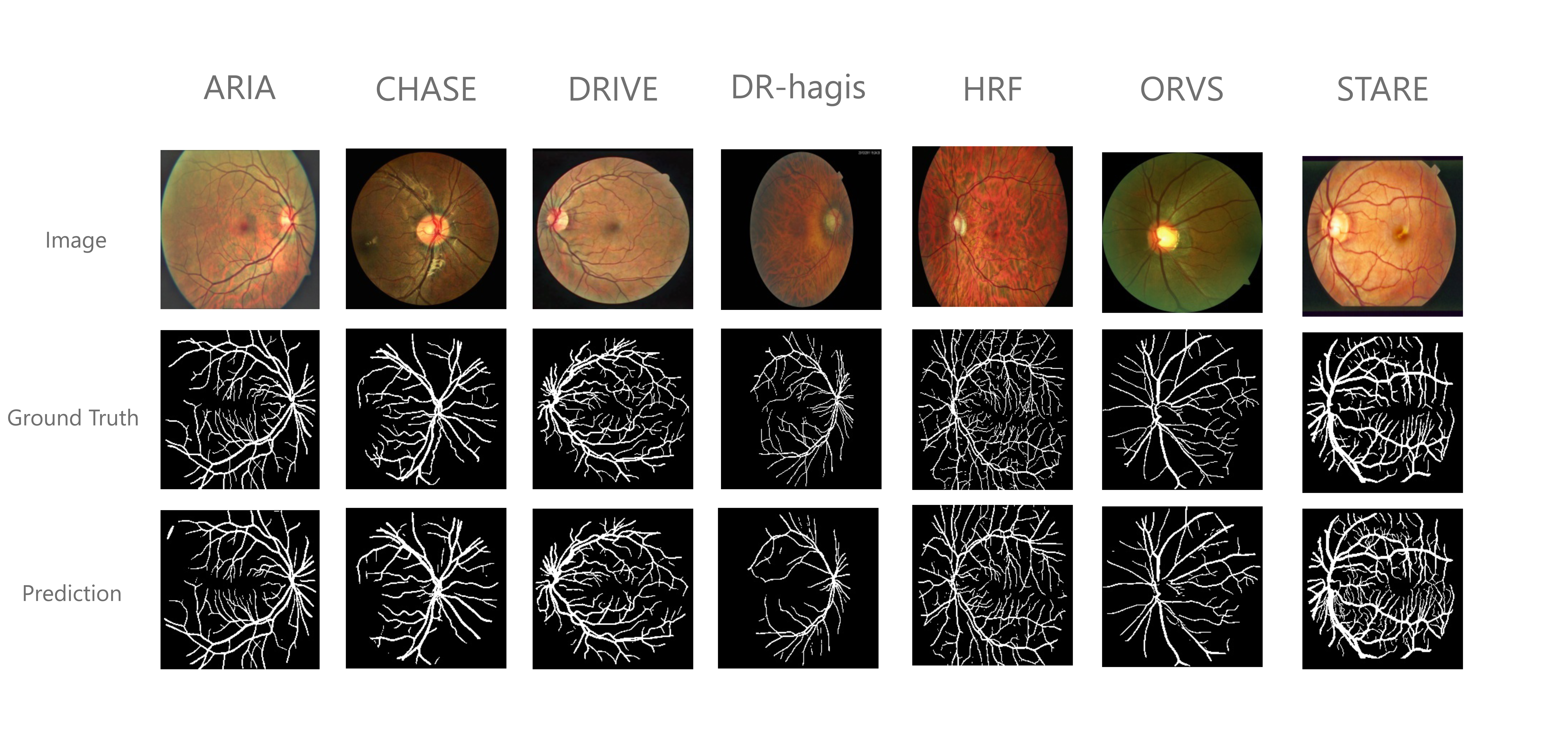}
\caption{Performance of our approach on images obtained from the datasets adopted in this study.}
\label{fig:VesselsDatasets}
\end{figure*}

\subsection{Implementation Details}

We implemented this model on a windows machine with a  NVIDIA GeForce 2060 RTX with 6 GB dedicated memory. We used the Python language to implement proposed approach using Keras with tensorflow back-end. 

Training was performed using the NAdam optimizer with learning rate set to 0.0001, $\beta_1$ = 0.9, $\beta_2$ = 0.999, $\epsilon= 10^{-8}$, and batch size of 2 images. During training, three callbacks were used. First, the model checkpoints would save the model whenever a smaller value was returned on validation data from the custom loss function when comparing to the value at the last checkpoint. Secondly, the learning rate was reduced by a factor of 0.5 whenever 25 epochs passed without any improvement in the validation loss values. Finally the training was stopped if 100 epochs passed without any improvement. When training our model we split the training and validation set by 85\% and 15\% respectively. These sets are selected randomly at each epoch. We used four evaluation methods to evaluate and compare our approach: namely, accuracy (Acc), dice coefficient (DC), sensitivity (Sen), specificity (Spec).

\subsection{Effectiveness of TL and IA }
To test the impact of using transfer learning (TL) and image augmentation (IA) when training our model, we conducted a series of experiments and then evaluated the model obtained by using the test images for all datasets. In this section, we show the overall performance without showing the performance with respect to each dataset. Note that, in all these experiments, we used the loss function defined in Eq.\ref{equ:FinalLoss}.

We first checked the performance of the model without transfer learning by randomly initializing weights using Gaussian distribution. Then we did an experiment using only rotated images with transfer learning. The third was using transfer learning and rotated and flipped images. The evaluation results for each of these experiments are shown in Table \ref{table:TL}. The results obtained show that using TL and IA together achieves the best results. Flipping helped in improving the results slightly from when using rotation only when augmenting images. Using only rotated images with transfer learning took around 319 epochs to converge while when using rotated and flipped images it needed 50 epochs.

\begin{table}[htb!]
\center
\caption {\label{table:TL} Performance comparison of proposed method with and without using transfer learning (TL) and/or image augmentation (IA). }
\begin{tabular}{lcccc}
\hline
Experiment              & \multicolumn{1}{c}{Acc} & \multicolumn{1}{c}{Sen} & \multicolumn{1}{c}{Spec} & \multicolumn{1}{c}{DC} \\\hline
Rotated+Flipped    & 93.98                   & 56.45                   & 98.98                    & 68.14                  \\
TL+RotatedOnly     & 95.58                   & 84.57                   & 96.60                    & 80.88                  \\
TL+Rotated+Flipped & 95.60                   & 85.18                   & 96.51                    & 80.98 \\\hline                
\end{tabular}
\end{table}

\subsection{Effectiveness of Loss Functions}
A well-known loss function for binary classification is the binary cross-entropy loss function. This loss function works well when the classes in the image are balanced. However, in our case, the object we are trying to segment represents 20\% or less of the image. Hence, we decided to use the Jaccard distance combined with BCE, as noted earlier.

We conducted three experiments to test which configuration would achieve the best results. First, we trained our model using the BCE loss function only, which is a built-in loss function in the Keras library; second, we trained it using the Jaccard distance-loss function only; and third, we trained it by combining both loss functions. The results obtained are shown in Table \ref{table:lossF}. We noted that there is a slight improvement in performance when we combine both loss functions compared to using either one of them alone. We also realized that using BCE and Jaccard distance together would help in converging faster, 50 epochs, than when using BCE with 71 epochs.

\begin{table}[htb!]
\center
\caption {\label{table:lossF} Performance of the model across different loss functions.}

\begin{tabular}{lcccc}
\hline
Loss Function        & Acc   & Sen   & Spec  & DC    \\ \hline
BCE                  & 95.25 & 86.58 & 95.92 & 80.23 \\ 
Jaccard Distance     & 95.21 & 85.24 & 96.12 & 79.82 \\ 
Jaccard Distance+BCE & 95.60 & 85.18 & 96.51 & 80.98 \\ \hline
\end{tabular}
\end{table}

\begin{table*}[htb!]
\center
\caption {\label{table:Retinal} Approaches for retinal vessels segmentation on the DRIVE and STARE datasets.} 
\begin{tabular}{lccccccccc}
\hline
Method & DRIVE    &             &            &                  &  & STARE    &             &            &                  \\ \cline{2-5} \cline{7-10} 
       & Acc & Sen & Spec & DC &  & ACC & Sen & Spec & DC \\ \hline
        \textbf{Unsupervised}      &          &             &            &                  &  &          &             &            &                  \\
 Nguyen et al.\cite{nguyen2013effective}      & 94.07         & -            &98.58            & 72.98                 &  &  93.24        &  -           & 98.63           &77.74                  \\

Roychowdhury, Sohini, et al. \cite{roychowdhury2015iterative}      & 94.9         &73.9             &97.8            &                  &  &  95.6        &73.2             &98.4            &                  \\

Memari et al. \cite{memari2019retinal}      & 96.1         &76.1             &98.1            &        -          &  &  95.1        &78.2            &96.5            &  -                \\

Zhao et al. \cite{zhao2014retinal}      & 94.7         &73.54             &97.89            &        -          &  &  95.09        &71.87            &97.67            & -                 \\

Khan et al. \cite{khan2016automatic}      & 95.1         &73.4             &96.7            &         -         &  &  95        &73.6            &97.1            & -                 \\

Zhang et al. \cite{zhang2016robust}      & 94.7         &74.3             &97.6            &         -         &  &  95.4        &76.7            &97.6            & -                 \\

Bankhead et al. \cite{bankhead2012fast}      & 93.7         &70.3             &97.1            &              -    &  &  93.2        &75.8            &95.0            & -                 \\

\textbf{Supervised }     &          &             &            &                  &  &          &             &            &                  \\

Wang et al.\cite{wang2019blood}       & 95.41         &76.48             &98.17            &80.93                  &  &96.40          &75.23             & 98.85           &81.25                  \\

Hu et al. \cite{hu2018retinal}       &95.33          & 77.72            &  97.93          &              -    &  &  96.32        &  75.43           & 98.14           &  -                \\
Oliveira et al. \cite{oliveira2018retinal}       & 95.76         &80.39             & 98.04           &-                  &  &96.94          & 83.15            & 98.58           & -                 \\
 Xia et al. \cite{xiao2018weighted}      & 96.55         & 77.15            & -           &                  -&  &  96.93        & 74.69            & -           &   -               \\
 Fu, Xu, Wong, et al. \cite{fu2016retinal}      & 95.20         & 76.00           &   -         &              -    &  & 95.80         & 74.10            &  -          & -                 \\
Yan et al. \cite{yan2018joint}       & 95.40         & 76.50            &  98.10          &          -        &  &96.10          & 75.80            &   97.50         &    -              \\
Brancati et al. \cite{brancati2017retinal}      &94.90          &78.20             &97.60           &  -                &  &   -     &        -     &        -    &   -               \\

Orlando et al. \cite{orlando2016discriminatively}      &-          &78.97             &96.84            &  78.41                &  &     -  & 76.80            &97.38            &   76.44               \\

Jin et al. \cite{jin2019dunet}      &95.66          &79.63             &98.00            &  82.37                &  & 96.41       & 75.95            &98.78            &   81.43               \\

 Proposed Method  &   95.61       & \textbf{82.67}            & 97.27           &   \textbf{82.45}              &&95.26  &  \textbf{85.61}        &96.57             &\textbf{84.06}                              \\
\hline
\end{tabular}
\end{table*}

\subsection{Comparison With Other Approaches}
To evaluate our proposed method, we compared it with approaches which had been tested on some of the same datasets we used, as shown in Table \ref{table:Retinal}. Unfortunately, these other approaches had not been evaluated with respect to all available datasets; hence, when comparing, we split our results per dataset so we could make a fair comparison.

We compared our approach to several studies, all of which used the Stare and DRIVE datasets (check Table \ref{table:Retinal}). Of the 16 approaches, only six used the CHASE  dataset (check Table\ref{table:Chase}). Two used the HRF (check table \ref{table:HRF}), and none used the DR-Hagis and the ARIA datasets (check table \ref{table:ORVS}). For each of these datasets, we show the performance of our model with respect to accuracy, sensitivity, specificity, and the dice coefficient. We also show, in Fig. \ref{fig:VesselsDatasets} the vessels segmented by our model for a sample image from each dataset along with ground truth and related retinal images. This figure clearly shows the variation in the orientations and resolutions of the vessels in each dataset.

\begin{table}[htb!]
\center
\caption {\label{table:Chase} Approaches for retinal vessels segmentation on CHASE dataset.} 
\begin{tabular}{lcccc}
\hline
Method           & Acc   & Sen   & Spec  & DC    \\\hline

Oliveira et al.\cite{oliveira2018retinal} & 96.53 & 77.79 & 98.64 & -     \\
Wang et al. \cite{wang2019blood}    & 96.03 & 77.30 & 97.92 & 78.09 \\
Memari et al. \cite{memari2019retinal}  & 93.90 & 73.80 & 96.80 & -     \\
Yan et al. \cite{yan2018joint}  & 96.10 & 76.33 & 98.09 & -     \\
Orlando et al. \cite{orlando2016discriminatively}  & - & 72.77 & 97.12 & 73.32    \\
Jin et al. \cite{jin2019dunet}  & 96.10 & 81.55 & 97.52 & 78.83    \\
Proposed Method & \textbf{96.83} & \textbf{90.21} & 97.34 & \textbf{85.46}\\\hline
\end{tabular}
\end{table}

\begin{table}[htb!]
\center
\caption {\label{table:HRF} Approaches for retinal vessels segmentation on HRF dataset.} 
\begin{tabular}{lcccc}
\hline
Method           & Acc   & Sen   & Spec  & DC    \\\hline

Orlando et al. \cite{orlando2016discriminatively}  & - & 78.74 & 95.84 & 71.58    \\
Jin et al. \cite{jin2019dunet}  & 96.51 & 74.64 & 98.74 & -    \\
Proposed Method & 95.07 & \textbf{91.56} & 95.09 & \textbf{81.90}\\\hline
\end{tabular}
\end{table}

\begin{table}[htb!]
\center
\caption {\label{table:ORVS} Performance of our approach on the DR-Hagis, ARIA, and ORVS datasets.} 
\begin{tabular}{lcccc}
\hline
Dataset           & Acc   & Sen   & Spec  & DC    \\\hline

DR-Hagis  & 96.87 & 67.13 & 98.57 & 71.49    \\
ARIA  & 95.31 & 81.94 & 96.40 & 77.59    \\
ORVS  & 96.52 & 84.32 & 97.19 & 78.11    \\
\hline
\end{tabular}
\end{table}

Our proposed approach achieved the highest sensitivity and dice coefficient among all approaches when tested against DRIVE and Stare. Moreover, our model achieved the highest accuracy, sensitivity, and dice coefficient on the CHASE dataset which had been tested against the same images. Furthermore, we achieved the highest accuracy, sensitivity, prediction, and Dice coefficient on the HRF dataset. We realized that, although some approaches performed slightly better than our method in terms of accuracy, we achieved better in terms of dice coefficient and sensitivity, which shows that our approach is more precised in segmenting the vessels than these approaches. For instance, in \cite{xiao2018weighted} researchers achieved the highest accuracy with less than 1\% difference from what was achieved by our model. However, our model outperformed their approach in terms of sensitivity, with more than a 5\% difference between our measurement and theirs. In \cite{jin2019dunet}, researchers achieved slightly better scores for accuracy and specificity, but our approach performed better in terms sensitivity and Dice coefficient. Given that the vessels occupy less than 20\% of the retinal image, accuracy alone is not good indicator for checking the performance of a model. Sensitivity and Dice coefficient are better indicators against which to evaluate a model when segmenting retinal vessels.

We also noticed that some approaches performed slightly better than ours in terms of accuracy and specificity with respect to some dataset(s) but that we outperformed those approaches when tested against other datasets. For instance, in \cite{oliveira2018retinal}, researchers achieved slightly better accuracy and precision than our approach when tested on the DRIVE datasets, but our model performed better when tested on the CHASE dataset.

Since none of the reported studies reported their performance on the DR-Hagis and ARIA datasets, we just reported the performance of our model with respect to these datasets by including our new public dataset, as shown in Table \ref{table:ORVS}. One reason the ARIA is not commonly used today is that it is no longer available online. However, it can still be retrieved by request, and we accordingly requested it. As for the DR-Hagis dataset, it is newly published. In general, our proposed approach shows superior performance with respect to the approaches reported in this study in terms of precisely segmenting vessels when exposed to images with various resolutions and orientations.

\section{Conclusion}
In this paper, we propose a deep learning based approach for vessels segmentation where we show the effectiveness of our customized model when being combined with transfer learning, image augmentation, and a customized loss function. Our approach achieve state of the art performance on vessels segmentation when compared to other modern approaches. We also contribute a new dataset the can be used by researchers for improving vessels segmentation. This will help researchers testing their approaches on images obtained from various sources with diverse data. Our model achieved an average accuracy, sensitivity, specificity, and dice coefficient of 95.60\%, 85.18\%,96.51\% and 80.98\% respectively.


\section*{Ethics}
The necessary ethics approval has been obtained by Health Research Ethics Board of Alberta (HREBA REB17-2302). No identifiable information has been shared and hence is not possible to identify corresponding individuals these images obtained from.





\bibliographystyle{IEEEtran}
\bibliography{IEEEexample.bib}

\begin{thebibliography}{10}
\providecommand{\url}[1]{#1}
\csname url@samestyle\endcsname
\providecommand{\newblock}{\relax}
\providecommand{\bibinfo}[2]{#2}
\providecommand{\BIBentrySTDinterwordspacing}{\spaceskip=0pt\relax}
\providecommand{\BIBentryALTinterwordstretchfactor}{4}
\providecommand{\BIBentryALTinterwordspacing}{\spaceskip=\fontdimen2\font plus
\BIBentryALTinterwordstretchfactor\fontdimen3\font minus
  \fontdimen4\font\relax}
\providecommand{\BIBforeignlanguage}[2]{{%
\expandafter\ifx\csname l@#1\endcsname\relax
\typeout{** WARNING: IEEEtran.bst: No hyphenation pattern has been}%
\typeout{** loaded for the language `#1'. Using the pattern for}%
\typeout{** the default language instead.}%
\else
\language=\csname l@#1\endcsname
\fi
#2}}
\providecommand{\BIBdecl}{\relax}
\BIBdecl

\bibitem{sarhan2019glaucoma}
A.~Sarhan, J.~Rokne, and R.~Alhajj, ``Glaucoma detection using image processing
  techniques: A literature review,'' \emph{Computerized Medical Imaging and
  Graphics}, p. 101657, 2019.

\bibitem{fu2017segmentation}
H.~Fu, Y.~Xu, S.~Lin, X.~Zhang, D.~W.~K. Wong, J.~Liu, A.~F. Frangi,
  M.~Baskaran, and T.~Aung, ``Segmentation and quantification for angle-closure
  glaucoma assessment in anterior segment oct,'' \emph{IEEE Transactions on
  Medical Imaging (TMI)}, pp. 1930 -- 1938, 2017.

\bibitem{issac2015adaptive}
A.~Issac, M.~P. Sarathi, and M.~K. Dutta, ``An adaptive threshold based image
  processing technique for improved glaucoma detection and classification,''
  \emph{Computer Methods and Programs in Biomedicine}, vol. 122, no.~2, pp.
  229--244, 2015.

\bibitem{szegedy2016rethinking}
C.~Szegedy, V.~Vanhoucke, S.~Ioffe, J.~Shlens, and Z.~Wojna, ``Rethinking the
  inception architecture for computer vision,'' in \emph{Proceedings of the
  IEEE Conference on Computer Vision and Pattern Recognition(CVPR)}, 2016, pp.
  2818--2826.

\bibitem{akbar2019automated}
S.~Akbar, M.~Sharif, M.~U. Akram, T.~Saba, T.~Mahmood, and M.~Kolivand,
  ``Automated techniques for blood vessels segmentation through fundus retinal
  images: A review,'' \emph{Microscopy Research and Technique}, vol.~82, no.~2,
  pp. 153--170, 2019.

\bibitem{mendonca2006segmentation}
A.~M. Mendonca and A.~Campilho, ``Segmentation of retinal blood vessels by
  combining the detection of centerlines and morphological reconstruction,''
  \emph{IEEE Transactions on Medical Imaging (TMI)}, vol.~25, no.~9, pp.
  1200--1213, 2006.

\bibitem{khan2016automatic}
M.~A. Khan, T.~A. Soomro, T.~M. Khan, D.~G. Bailey, J.~Gao, and N.~Mir,
  ``Automatic retinal vessel extraction algorithm based on contrast-sensitive
  schemes,'' in \emph{2016 International Conference on Image and Vision
  Computing New Zealand (IVCNZ)}.\hskip 1em plus 0.5em minus 0.4em\relax IEEE,
  2016, pp. 1--5.

\bibitem{bankhead2012fast}
P.~Bankhead, C.~N. Scholfield, J.~G. McGeown, and T.~M. Curtis, ``Fast retinal
  vessel detection and measurement using wavelets and edge location
  refinement,'' \emph{PloS One}, vol.~7, no.~3, p. e32435, 2012.

\bibitem{nguyen2013effective}
U.~T. Nguyen, A.~Bhuiyan, L.~A. Park, and K.~Ramamohanarao, ``An effective
  retinal blood vessel segmentation method using multi-scale line detection,''
  \emph{Pattern Recognition}, vol.~46, no.~3, pp. 703--715, 2013.

\bibitem{roychowdhury2015iterative}
S.~Roychowdhury, D.~D. Koozekanani, and K.~K. Parhi, ``Iterative vessel
  segmentation of fundus images,'' \emph{IEEE Transactions on Biomedical
  Engineering (TBME)}, vol.~62, no.~7, pp. 1738--1749, 2015.

\bibitem{zhao2014retinal}
Y.~Q. Zhao, X.~H. Wang, X.~F. Wang, and F.~Y. Shih, ``Retinal vessels
  segmentation based on level set and region growing,'' \emph{Pattern
  Recognition}, vol.~47, no.~7, pp. 2437--2446, 2014.

\bibitem{memari2019retinal}
N.~Memari, A.~R. Ramli, M.~I.~B. Saripan, S.~Mashohor, and M.~Moghbel,
  ``Retinal blood vessel segmentation by using matched filtering and fuzzy
  c-means clustering with integrated level set method for diabetic retinopathy
  assessment,'' \emph{Journal of Medical and Biological Engineering}, vol.~39,
  no.~5, pp. 713--731, 2019.

\bibitem{shen2017deep}
D.~Shen, G.~Wu, and H.-I. Suk, ``Deep learning in medical image analysis,''
  \emph{Annual Review of Biomedical Engineering}, vol.~19, pp. 221--248, 2017.

\bibitem{xiao2018weighted}
X.~Xiao, S.~Lian, Z.~Luo, and S.~Li, ``Weighted res-unet for high-quality
  retina vessel segmentation,'' in \emph{2018 9th International Conference on
  Information Technology in Medicine and Education (ITME)}.\hskip 1em plus
  0.5em minus 0.4em\relax IEEE, 2018, pp. 327--331.

\bibitem{oliveira2018retinal}
A.~Oliveira, S.~Pereira, and C.~A. Silva, ``Retinal vessel segmentation based
  on fully convolutional neural networks,'' \emph{Expert Systems with
  Applications}, vol. 112, pp. 229--242, 2018.

\bibitem{hu2018retinal}
K.~Hu, Z.~Zhang, X.~Niu, Y.~Zhang, C.~Cao, F.~Xiao, and X.~Gao, ``Retinal
  vessel segmentation of color fundus images using multiscale convolutional
  neural network with an improved cross-entropy loss function,''
  \emph{Neurocomputing}, vol. 309, pp. 179--191, 2018.

\bibitem{ronneberger2015u}
O.~Ronneberger, P.~Fischer, and T.~Brox, ``U-net: Convolutional networks for
  biomedical image segmentation,'' in \emph{International Conference on Medical
  Image Computing and Computer-Assisted Intervention (MICCAI)}.\hskip 1em plus
  0.5em minus 0.4em\relax Springer, 2015, pp. 234--241.

\bibitem{pizer1987adaptive}
S.~M. Pizer, E.~P. Amburn, J.~D. Austin, R.~Cromartie, A.~Geselowitz, T.~Greer,
  B.~ter Haar~Romeny, J.~B. Zimmerman, and K.~Zuiderveld, ``Adaptive histogram
  equalization and its variations,'' \emph{Computer Vision, Graphics, and Image
  Processing}, vol.~39, no.~3, pp. 355--368, 1987.

\bibitem{wu2018group}
Y.~Wu and K.~He, ``Group normalization,'' in \emph{Proceedings of the European
  Conference on Computer Vision (ECCV)}, 2018, pp. 3--19.

\bibitem{pan2009survey}
S.~J. Pan and Q.~Yang, ``A survey on transfer learning,'' \emph{IEEE
  Transactions on Knowledge and Data Engineering}, vol.~22, no.~10, pp.
  1345--1359, 2009.

\bibitem{fraz2012ensemble}
M.~M. Fraz, P.~Remagnino, A.~Hoppe, B.~Uyyanonvara, A.~R. Rudnicka, C.~G. Owen,
  and S.~A. Barman, ``An ensemble classification-based approach applied to
  retinal blood vessel segmentation,'' \emph{IEEE Transactions on Biomedical
  Engineering}, vol.~59, no.~9, pp. 2538--2548, 2012.

\bibitem{holm2017dr}
S.~Holm, G.~Russell, V.~Nourrit, and N.~McLoughlin, ``Dr hagis—a fundus image
  database for the automatic extraction of retinal surface vessels from
  diabetic patients,'' \emph{Journal of Medical Imaging}, vol.~4, no.~1, p.
  014503, 2017.

\bibitem{niemeijer2004comparative}
M.~Niemeijer, J.~Staal, B.~van Ginneken, M.~Loog, and M.~D. Abramoff,
  ``Comparative study of retinal vessel segmentation methods on a new publicly
  available database,'' in \emph{Medical Imaging 2004: Image Processing}, vol.
  5370.\hskip 1em plus 0.5em minus 0.4em\relax International Society for Optics
  and Photonics, 2004, pp. 648--656.

\bibitem{odstrcilik2013retinal}
J.~Odstrcilik, R.~Kolar, A.~Budai, J.~Hornegger, J.~Jan, J.~Gazarek, T.~Kubena,
  P.~Cernosek, O.~Svoboda, and E.~Angelopoulou, ``Retinal vessel segmentation
  by improved matched filtering: evaluation on a new high-resolution fundus
  image database,'' \emph{IET Image Processing}, vol.~7, no.~4, pp. 373--383,
  2013.

\bibitem{hoover2000locating}
A.~Hoover, V.~Kouznetsova, and M.~Goldbaum, ``Locating blood vessels in retinal
  images by piecewise threshold probing of a matched filter response,''
  \emph{IEEE Transactions on Medical Imaging (TMI)}, vol.~19, no.~3, pp.
  203--210, 2000.

\bibitem{yan2018joint}
Z.~Yan, X.~Yang, and K.-T. Cheng, ``Joint segment-level and pixel-wise losses
  for deep learning based retinal vessel segmentation,'' \emph{IEEE
  Transactions on Biomedical Engineering}, vol.~65, no.~9, pp. 1912--1923,
  2018.

\bibitem{wang2019blood}
X.~Wang, X.~Jiang, and J.~Ren, ``Blood vessel segmentation from fundus image by
  a cascade classification framework,'' \emph{Pattern Recognition}, vol.~88,
  pp. 331--341, 2019.

\bibitem{orlando2016discriminatively}
J.~I. Orlando, E.~Prokofyeva, and M.~B. Blaschko, ``A discriminatively trained
  fully connected conditional random field model for blood vessel segmentation
  in fundus images,'' \emph{IEEE Transactions on Biomedical Engineering},
  vol.~64, no.~1, pp. 16--27, 2016.

\bibitem{zhang2016robust}
J.~Zhang, B.~Dashtbozorg, E.~Bekkers, J.~P. Pluim, R.~Duits, and B.~M. ter
  Haar~Romeny, ``Robust retinal vessel segmentation via locally adaptive
  derivative frames in orientation scores,'' \emph{IEEE Transactions on Medical
  Imaging (TMI)}, vol.~35, no.~12, pp. 2631--2644, 2016.

\bibitem{fu2016retinal}
H.~Fu, Y.~Xu, D.~W.~K. Wong, and J.~Liu, ``Retinal vessel segmentation via deep
  learning network and fully-connected conditional random fields,'' in
  \emph{2016 IEEE 13th International Symposium on Biomedical Imaging
  (ISBI)}.\hskip 1em plus 0.5em minus 0.4em\relax IEEE, 2016, pp. 698--701.

\bibitem{brancati2017retinal}
N.~Brancati, M.~Frucci, D.~Gragnaniello, and D.~Riccio, ``Retinal vessels
  segmentation based on a convolutional neural network,'' in
  \emph{Iberoamerican Congress on Pattern Recognition}.\hskip 1em plus 0.5em
  minus 0.4em\relax Springer, 2017, pp. 119--126.

\bibitem{jin2019dunet}
Q.~Jin, Z.~Meng, T.~D. Pham, Q.~Chen, L.~Wei, and R.~Su, ``Dunet: A deformable
  network for retinal vessel segmentation,'' \emph{Knowledge-Based Systems},
  vol. 178, pp. 149--162, 2019.

\end{thebibliography}
%



\end{document}